\documentclass[%
 reprint,
 amsmath,amssymb,
 aps,
 prl,
longbibliography
]{revtex4-1}

\usepackage{graphicx}
\usepackage{bm,braket}
\usepackage{hyperref}
\usepackage[utf8]{inputenc} 
\usepackage[T1]{fontenc} 

\usepackage{pgffor}
\usepackage{pdfpages}
\makeatletter
\AtBeginDocument{\let\LS@rot\@undefined}
\makeatother

\begin{document}

\title{Self-Pinning Transition of a Tonks-Girardeau Gas in a Bose-Einstein Condensate}

\author{Tim Keller}
\email{tim.keller@oist.jp}
\author{Thom\'as Fogarty}
\author{Thomas Busch}
\affiliation{Quantum Systems Unit, Okinawa Institute of Science and Technology Graduate University, Onna-son, Okinawa 904-0495, Japan}

\date{\today}

\begin{abstract}
We show that a Tonks-Girardeau (TG) gas that is immersed in a Bose-Einstein condensate can undergo a transition to a crystal-like Mott state with regular spacing between the atoms without any externally imposed lattice potential. We characterize this phase transition as a function of the interspecies interaction and temperature of the TG gas, and show how it can be measured via accessible observables in cold atom experiments. We also develop an effective model that accurately describes the system in the pinned insulator state and which allows us to derive the critical temperature of the transition.
\end{abstract}

\maketitle

\textit{Introduction.} --- Quantum phase transitions are a hallmark of quantum many-body physics and responsible for systems being able to access new states and obtain unique properties \cite{Sachdev:2011}. In lattice systems of cold atoms at effectively zero temperature the celebrated superfluid to Mott-insulator transition was first experimentally observed by Greiner {\it et al.} \cite{Greiner2002} and sparked a large effort in observing condensed matter physics in these highly controllable systems \cite{Ahufinger:2012}. More recently, the possibilities to reach new parameter regimes have also led to these systems becoming highly successful quantum simulators \cite{Gross:17}.

In one dimension a commensurate-incommensurate phase transition emerges in the regime of strong repulsive interactions, the well-known Tonks-Girardeau (TG) limit \cite{girardeau1960relationship}, where ordering of particles in a lattice potential is governed by the filling ratio. In this limit the transition takes place even for infinitesimal lattice depths \cite{buchler2003commensurate} and it has been experimentally observed by Haller {\it et al.} \cite{haller2010pinning}. Since one-dimensional systems in the TG limit are highly analytically accessible even for larger numbers of particles, this system has received a lot of attention in recent years, especially with respect to its out-of-equilibrium dynamics \cite{lelas2012pinning,Cartarius2015,astrakharchik2016one,boeris2016mott,atas2017exact,Atas2017b,mikkelsen2018static,Atas2019,Atas2020,Fogarty_2020}.

While these quantum phase transitions are induced by the control of static external fields, more complex phenomena can be explored in systems where the particles exert a backaction on their environments. For example, long-range interactions can be created between particles confined in an optical cavity, leading to self-organization phase transitions such as the Dicke transition \cite{Baumann2010,fernandez2010quantum,habibian2013bose,schutz2015thermodynamics} and the superfluid-supersolid transition \cite{Mottl1570}. Analogous systems are two-component Bose-Einstein condensates (BECs) where competition between the interspecies and intraspecies interactions can realize miscible-immiscible transitions, where the two components avoid spatial overlap to reduce interaction energies \cite{Myatt1997,Hall1998,Mertes2007,Papp2008,Tojo2010}, as well as Bose-Fermi mixtures \cite{miyakawa2004peierls}. Beyond the mean-field regime, phase separation and composite fermionization have been explored in true many-body systems, allowing one to probe correlations and nonequilibrium dynamics from systems of few \cite{Zollner2008,Pflanzer2010,MAGM2013,MAGM2013b,MAGM2014,MAGM_2014,Pecak2016,Pyzh_2018,Pyzh2020,Mistakidis2020,barfknecht2021generation} to many particles \cite{Mistakidis_2018,Bougas2021}.

In this work we consider a system where a small number of strongly interacting atoms in the TG limit is immersed in a much larger, but weakly correlated, background BEC \cite{MAGM2013}. Here, interspecies interactions can create an effective mean-field potential for the TG gas, which is highly nonlinear since it depends on the positions of the individual TG atoms. In analogy to the pinning transition, we show that for suitably strong interspecies interaction the TG gas self-organizes into a regular structure thereby creating its own perfectly commensurate mean-field potential in the BEC.
Using an effective description in terms of nonlinear localized single particle states for the TG atoms, we are able to find an analytical expression for the energy gap that opens at the phase transition point as a function of the interaction strength between both components. Finally, we describe this self-pinning transition at finite temperature and derive an expression for the critical temperature below which the pinned state can emerge and discuss how it can be observed experimentally.

\textit{Model.} --- We consider a highly anisotropic cigar-shaped BEC of $N_c$ particles with the radial degrees of freedom restricted to their respective ground states, leading to an effectively one-dimensional setting, which in the mean-field limit is described by a macroscopic wave function $\psi (x)$. Into the condensate a small sample of $N$ particles is immersed, which is described by a full many-particle wave function, $\Phi({\bf x}=x_1,x_2,\dots, x_{N})$. At low temperatures all interactions can be approximated by pointlike pseudopotentials and quantified by scattering lengths only. If we assume that the couplings between the immersed atoms and the ones in the BEC are weak, the interactions between the two components can be described by a straightforward density coupling, leading to the coupled evolution equations
\begin{align}
    \label{eq:GPE}
    i\dot\psi(x)=&\left[-\frac{1}{2}\frac{\partial^2}{\partial x^2}+g_m|\Phi|^2+g_c|\psi|^2\right]\psi(x)\\
    \label{eq:MB}
    i\dot\Phi({\bf x})=&\left[\sum_{l}-\frac{1}{2}\frac{\partial^2}{\partial x_l^2} +V(x_l)+g_m|\psi|^2+V_\text{int}\right]\Phi({\bf x}),
\end{align}
where for simplicity we have set $\hbar$ and all masses equal to one. The interactions between the immersed atoms are described by $V_\text{int}=g\sum_{k<l}^{N}\delta(|x_k-x_l|)$, and $g$, $g_c$, and $g_m$ are proportional to the scattering lengths describing the interaction strengths within the immersed gas, within the condensate and between the two components, respectively. We assume that the condensate is in free space with an average density $n_c\equiv N_c/L_c = \mu_0/g_c$, whereas the immersed component sees a box potential of width $L$ with $V(x)\equiv 0$ for $|x| \leq L/2$ and $V(x)\equiv\infty$ otherwise.

While Eq.~\eqref{eq:MB} for the immersed component is hard to solve for larger particle numbers and for arbitrary values of $g$, the TG limit of strong interactions ($g\rightarrow\infty$) allows for exact solutions due to the Bose-Fermi mapping theorem \cite{girardeau1960relationship,yukalov2005fermi}. In this limit the interaction term can be replaced by a boundary condition resembling the Pauli exclusion principle, and one can solve a system of noninteracting fermionic particles, while making sure that the bosonic symmetry is maintained. This means that all that is required is the knowledge of the single particle eigenstates $\phi_n(x)$ with eigenenergies $E_n$ which are the solutions of Eq.~\eqref{eq:MB} with $V_\text{int}=0$. The density of the TG gas at zero temperature, which is the quantity to which the BEC couples in Eq.~\eqref{eq:GPE}, is then simply given by
\begin{equation}
\rho(x)=|\Phi(x)|^2 = \sum_{n=1}^{N}\left|\phi_n(x)\right|^2 \, .
\end{equation}
In the following we will concentrate on the TG regime for the immersed component.

\begin{figure}
    \centering
    \includegraphics[width=\columnwidth]{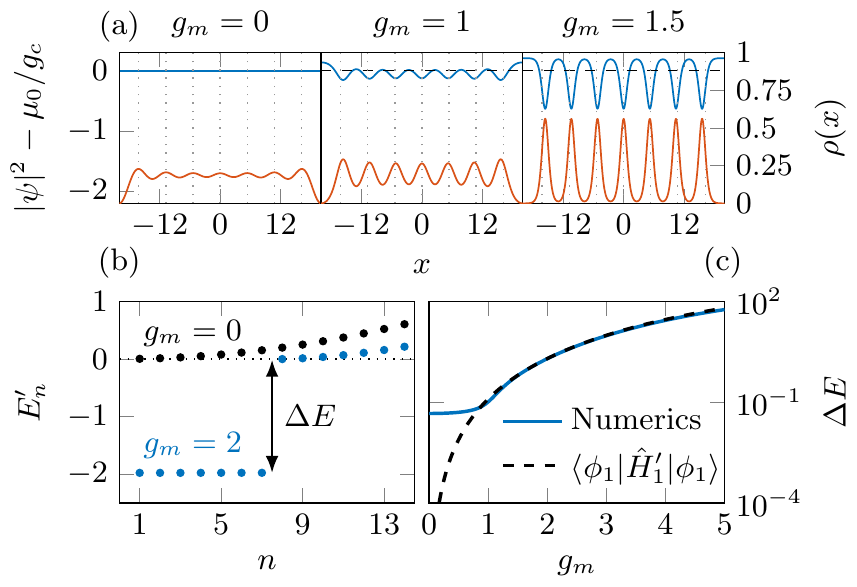}
\caption{(a) TG gas density $\rho(x)$ (red lines) and BEC depletion $|\psi|^2 - \mu_0/g_c$ (blue lines) for $N=7$ TG atoms and increasing interaction strengths $g_m$. The dotted lines indicate the maxima of the TG density in the noninteracting case $g_m=0$.  (b) Spectrum of the TG gas $E_n' = E_n-g_m\tilde{\mu}/g_c$ for $g_m=0$ (superfluid phase, black dots) and $g_m=2$ (pinned phase, blue dots). (c) Size of the energy gap as a function of $g_m$. The black dashed line shows the approximation $\Delta E \approx \langle \phi_1|\hat{H}'_1|\phi_1\rangle = |a_\mathrm{pin}^2/6-2a_0a_\mathrm{pin}/3|$. Other parameters are $g_c=1$ and $\mu_0=200$.}
\label{fig:zero_temp}
\end{figure}

\textit{Pinned states.} --- Solving the coupled evolution equations numerically, one can identify three different regimes as a function of the interaction strength between the TG atoms and the BEC, which are depicted in Fig.~\ref{fig:zero_temp}(a). If the two components do not interact, the immersed atoms are delocalized over the full box  and the condensate density is flat. However, for finite  values of $g_m$ the atoms start to localize in a regularly spaced pattern, while at the same time creating their own matter wave lattice potential in the BEC. This trend continues for increasing $g_m$, until the overlap between neighboring TG atoms becomes zero.

This behavior is quite different from the well-known miscible-immiscible phase transition one would expect in repulsively interacting multicomponent systems \cite{pethick2008bose}. However, it is strongly reminiscent of the pinning phase transition known to occur for a single-component TG gas \cite{buchler2003commensurate,haller2010pinning}, where the individual atoms become localized at individual lattice sites irrespective of the lattice depth. While in our situation no external lattice is applied, the interaction between the two components leads to an arrangement of the BEC density into a periodic pattern that resembles a standing matter wave which pins the atoms into the mean-field potential minima. The numerically obtained energy spectrum of the TG gas [in the shifted reference frame of the effective Hamiltonian in Eq.~\eqref{eq:tonksham_prime}, see below] is shown in Fig.~\ref{fig:zero_temp}(b) and one can see that in the pinned regime it exhibits a characteristic gap. The size of this gap, $\Delta E=E_{N+1}-E_{N}$, increases with increasing intercomponent interaction strength [see  Fig.~\ref{fig:zero_temp}(c)], therefore signaling the presence of an insulating phase in which the individual particles become localized in the matter-wave potential. For values below $g_m\approx 1$ the gap closes and the TG density is delocalized, therefore possessing superfluid properties. This is due to the finite size of our system, and the exact value of $g_m$ for which this happens depends on the number of particles $N$ as detailed in the Supplemental Material \cite{suppmat}.

To determine the size of the gap analytically, let us concentrate on the regime where the TG gas atoms are tightly localized. The overlap between adjacent particles then vanishes and one can use an effective single particle description for the immersed system. If the intercomponent interaction is small, $g_m\ll\mu_0 L/N$ , the BEC can be considered to be in the Thomas-Fermi limit, as the deviations of the density from the constant solution for $g_m=0$ are only small. This allows one to neglect the kinetic energy in Eq.~\eqref{eq:GPE} and leads to
\begin{equation}
    \psi(x,t) = \sqrt{\frac{1}{g_c}\left(\tilde{\mu}-g_m|\Phi|^2\right)}
            e^{-i\tilde{\mu} t}.
    \label{eq:tf_approximation}
\end{equation}
Here the modified chemical potential $\tilde{\mu} = \mu_0\left(1 + \frac{g_mN}{g_cN_c}\right)$ accounts for the change in condensate density due to the interaction with the immersed atoms. With this the relevant Hamiltonian for a single atom can be written as \cite{bruderer2008self,blinova2013two,grusdt2017bose,mistakidis2019quench,will2021polaron,schmidt2021self}
\begin{equation}
    \hat{H}_1' = -\frac{1}{2}\frac{\partial^2}{\partial x^2} - \frac{g_m^2}{g_c}\left|\phi_1(x)\right|^2,
    \label{eq:tonksham_prime}
\end{equation}
where the ground state energy is shifted by a constant term $E_{1}' = E_{1} - \frac{g_m}{g_c}\tilde{\mu}$. This Hamiltonian has a well-known nonlinear structure that allows for solitonlike localized solutions of inverse width $a_0$ of the form \cite{kevrekidis2007emergent}
\begin{equation}
    \phi_1(x) = \sqrt{\frac{a_0}{2}}\frac{1}{\cosh\left(a_0x\right)} \quad\text{ with }\quad a_0 = \frac{g_m^2}{2g_c}
    \label{eq:bright_soliton}
\end{equation}
and $E_{1}' = -a_0^2/2$.
However, the Thomas-Fermi limit does not take into account the energies that are needed by the immersed atom to displace the BEC density and by the BEC to keep the single atom from dispersing. Considering this will lead to a reduction in the peak height of the wave function of the immersed atom, $a_\mathrm{pin}<a_0$, and to a reduction in the density dip appearing in the BEC. For moderate interaction strengths $g_m$, the width of the atomic wave function and the density dip are proportional to each other and it is possible to find a closed expression $a_\mathrm{pin}= a_0\epsilon^{-1}(\sqrt{1+2\epsilon} - 1)$ with $\epsilon = 6a_0^2/5\tilde{\mu}$ (see Supplemental Material \cite{suppmat}).
The total energy of the coupled system of the BEC and the full TG gas in this pinned state is then given by
\begin{equation}
    E_\mathrm{pin} =  N\left( \frac{g_m^2}{30\tilde{\mu}g_c}a_\mathrm{pin}^3+ \frac{a_\mathrm{pin}^2}{6} - \frac{g_m^2}{6g_c} a_\mathrm{pin}\right) +\frac{\tilde{\mu}^2L_c}{2g_c} \, ,
\label{eq:energy_pinned}
\end{equation}
where the expression in the parentheses is the energy of a single atom.
In fact, atoms with energy $E_n'<0$ are pinned within the mean-field potential, while states with energy $E_n'\geq 0$ are delocalized over the whole system [see  Fig.~\ref{fig:zero_temp}(b)]. In the pinned phase one can therefore think of the TG atoms as being in Wannier-type states, which all have the same energy that lies slightly above the expected ground state energy $E'_{1} = -a_0^2/2$ of the effective Hamiltonian in Eq.~\eqref{eq:tonksham_prime}. By using the modified inverse width $a_\mathrm{pin}$ for the ground state, $\phi_1(x)=\sqrt{a_\mathrm{pin}/2}\cosh^{-1}(a_\mathrm{pin} x)$, an approximation to the numerically observed single particle energies can be given as $E'_{1}\approx \langle \phi_1|\hat{H}'_1|\phi_1\rangle = \frac{a_\mathrm{pin}^2}{6} - \frac{2}{3}a_0a_\mathrm{pin}$ \cite{suppmat}.
Since on the shifted energy scale $E'_{N+1}\approx 0$ (if the box is large enough), one can directly write an approximate expression for the energy gap as $\Delta E = E'_{N+1} - E'_{N} \approx |E_1'|$. This closely matches the numerical solutions for $g_m>1$ [see Fig.~\ref{fig:zero_temp}(c)]; however, it becomes unreliable in the regime of weak interactions, $g_m<1$, where the effective single particle description breaks down. Here the TG atoms start to overlap and their energy spectrum is dominated by the inherently strong interactions; therefore, a full many-body description is required. It is important to note that the scaling of the energy gap as $\Delta E \sim a_0^2\sim g_m^4/g_c^2$ is in line with the linear scaling $\Delta E \sim V_0$ reported for the pinning transition in an external lattice potential of strength $V_0$ \cite{buchler2003commensurate,haller2010pinning}, if in equilibrium the nonlinearity in the effective Hamiltonian Eq.~\eqref{eq:tonksham_prime} is regarded as an external potential $V(x)\sim V_0/\cosh^2(a_\mathrm{pin}x)$ with $V_0=a_0^2$.

Since deep in the pinned state the overlap between different atoms is zero, the TG gas density can simply be written as an arrangement of single impurity densities
\begin{equation}
    \rho_\mathrm{pin}(x) = \frac{a_\mathrm{pin}}{2}\sum_{n=1}^{N}\frac{1}{\cosh^2\left[a_\mathrm{pin}\left(x-x_n\right)\right]},
    \label{eq:rho_pinned}
\end{equation}
at positions $x_n$ with inverse width $a_\mathrm{pin}$.
If we consider the pinned state to be the result of an adiabatic ramp from $g_m=0$ to some final value $g_m>0$, these positions are approximately given by the maxima of their initial density in the infinite box $\rho(g_m=0,x)=\frac{2}{L}\sum_{n=1}^{N}\sin^2\left[\frac{n\pi}{L}\left(x+\frac{L}{2}\right)\right]$ which are determined by the odd solutions of $(2N+1)\tan(\pi z) = \tan((2N + 1)\pi z)$ for $0\leq z \leq 1$. This is highlighted in Fig.~\ref{fig:zero_temp}(a) where one can see that the positions of the tightly localized particles $x_n$ are equally spaced on the approximate order of $L/N$ when $g_m=1.5$.

\begin{figure}
    \centering
    \includegraphics[width=\columnwidth]{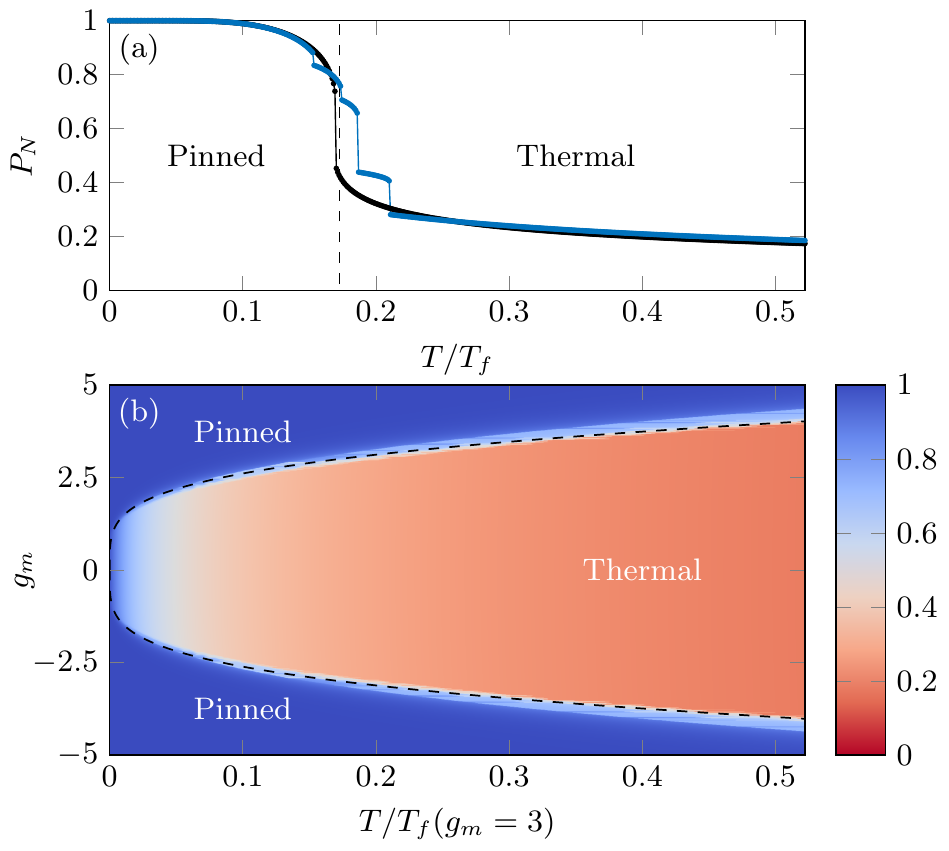}
    \caption{(a) Average ground state occupancy $P_{N}=\sum_{n=1}^{N}f_n/N$ for the system of $N = 7$ particles at $g_m=3$. The black line indicates the analytically determined value $f_\mathrm{pin}$ according to Eq. \eqref{eq:occupancy_criterion} and the black dashed line indicates the critical temperature $T_\mathrm{crit} = 0.1726\,T_f$ according to Eq. \eqref{eq:Tcrit}. (b) Phase diagram as a function of $g_m$ and temperature $T$. The color coding shows the ground state occupancy $P_{N}$. The black dashed line shows the analytically determined $T_\mathrm{crit}$ according to Eq. \eqref{eq:Tcrit}. The temperature scaling has been chosen for fixed $T_f(g_m=3)$ to be consistent with the other figures. Other parameters are $g_c=1$ and $\mu_0=200$.}
    \label{fig:finite_temp}
\end{figure}

\textit{Finite temperature.} --- While at zero temperature the most obvious manifestation of the self-organized pinning transition is a vanishing overlap between the different TG atoms, at finite temperatures
this tight localization of the single particle states is not guaranteed due to the presence of thermal excitations.
In this case the TG gas density $\rho(x)$ is given by the diagonal of the reduced single particle density matrix (RSPDM)
\begin{eqnarray}
    \rho(x,x')&=&\frac{1}{\mathcal{Z}}\sum_n e^{-\beta(E_n^N-\mu N)} \int dx_2 \dots dx_N  \nonumber \\
    && \times \Phi_n(x,x_2,\dots,x_N) \Phi_n^{*}(x',x_2,\dots,x_N).
    \label{eq:rspdm}
\end{eqnarray}
Here $\beta = 1/k_\mathrm{B}T$, $\mu$ is the chemical potential and $\mathcal{Z}=\sum_n e^{-\beta(E_n^N-\mu N)}$ is the grand-canonical partition function with $E_n^{N}$ being the total energy of the many-body wave function $\Phi_n$. While calculating Eq.~\eqref{eq:rspdm} is not an easy task, it was shown in Ref. \cite{Lenard1966} that the RSPDM of the TG gas can be written in terms of the RSPDM of spinless fermions.
More recently, efficient techniques to calculate Eq.~\eqref{eq:rspdm} for the finite temperature TG gas using just the single particle states $\phi_n(x)$ were presented \cite{atas2017exact,Atas2017b,Ovidiu2020}. Through this mapping the density of the TG gas can be written as $\rho(x) = \sum_{n=1}^\infty f_n \left|\phi_n(x)\right|^2$ where $f_n = \lbrace\exp\left[\beta\left(E_n - \mu\right)\right] + 1\rbrace^{-1}$ is the Fermi-Dirac distribution and $\mu$ is fixed by the number of atoms $N = \sum_{n=1}^\infty f_n$. Note that due to the large difference in particle number, we will assume that the BEC is still effectively at zero temperature.

To include the effect of temperature into the single impurity model one can replace $\phi_1\rightarrow \sqrt{f_1}\phi_1$ in the  Hamiltonian \eqref{eq:tonksham_prime}, which corresponds to an effective reduction of the interaction between the BEC and the TG gas. However, this results in a changed energy $E_1'$ which, in turn, leads to a modified occupancy $f_1$, etc..  The resulting state therefore needs to be determined from the self-consistency criterion
\begin{equation}
    f_\mathrm{pin} = \frac{1}{\exp\left\lbrace\beta\left[E(f_\mathrm{pin}) - \mu(f_\mathrm{pin})\right]\right\rbrace + 1} \;,
    \label{eq:occupancy_criterion}
\end{equation}
where the energy is given by the ground state energy of the  single-impurity Hamiltonian $\hat{H}'$ of Eq.~\eqref{eq:tonksham_prime} as $E(f_\mathrm{pin}) = E_1'\approx \frac{a_\mathrm{pin}^2}{6} - \frac{2}{3}f_\mathrm{pin}a_0a_\mathrm{pin}$.
The reduced peak height $a_{\mathrm{pin}}$ also depends on the Fermi-Dirac factors since it is calculated for $g_m^\mathrm{eff} =g_m\sqrt{f_\mathrm{pin}}$. The remaining factors $f_n$ with $n\geq N+1$, which correspond to nonpinned states, are required to determine the chemical potential $\mu(f_\mathrm{pin})$. Since these states are not trapped by the mean-field potential and therefore exist in the continuum, their energies can be well approximated by the energy spectrum of the box potential plus an energy offset given by the average density of the BEC.

The average occupation of the $N$ lowest states $P_{N}=\sum_{n=1}^{N}f_n/N$, which is equal to one in the pinned state at $T=0$, is shown in Fig.~\ref{fig:finite_temp}(a). Here and in the following we give the temperature in units of $T_f=\Delta E/k_B$ where $\Delta E$ is the average energy of the band gap in the pinned state at $T=0$. One can see that in a narrow temperature band sudden jumps occur in the probability indicating the ejection of particles from the pinned phase.
For large temperatures further discontinuities are absent and the ground state occupation takes values $P_{N}\ll 1$ implying that the many-body state is strongly delocalized. One can therefore use the quantity $P_N$ to map the phase diagram of this pinning transition as a function of $g_m$ and $T$, see Fig.~\ref{fig:finite_temp}(b). As one would expect, pinned states with larger intercomponent interactions are more robust to the effects of temperature, as the energy gap protects the ground state from thermal excitations. While the occupancy of the ground state allows us to determine both the pinned and delocalized TG phases, other order parameters show similar results, i.e.~energy and coherence functions (not shown).

The critical temperature for the pinning transition can also be determined from Eq.~\eqref{eq:occupancy_criterion} by looking for the point where the change in $f_\text{pin}$ is maximal, i.e.~where the left- and right-hand sides of Eq.~\eqref{eq:occupancy_criterion} are tangent to each other.
This gives \cite{suppmat}
\begin{equation}
    \frac{T_\mathrm{crit}}{T_f} = C(f^*)\frac{\sqrt{1+2\left(f^*\right)^2\epsilon}-1}{\sqrt{1+2\epsilon}-1} \, ,
    \label{eq:Tcrit}
\end{equation}
where $C(f^*)$ is a numerical constant and $f^*\approx 2/3$ is the value of $f_\mathrm{pin}$ at the critical temperature. The critical temperature is indicated in both plots in Fig.~\ref{fig:finite_temp} as a black dashed line and it can be seen to be in good agreement with the transition region observed in the numerical simulations.

\begin{figure}
    \centering
    \includegraphics[width=\columnwidth]{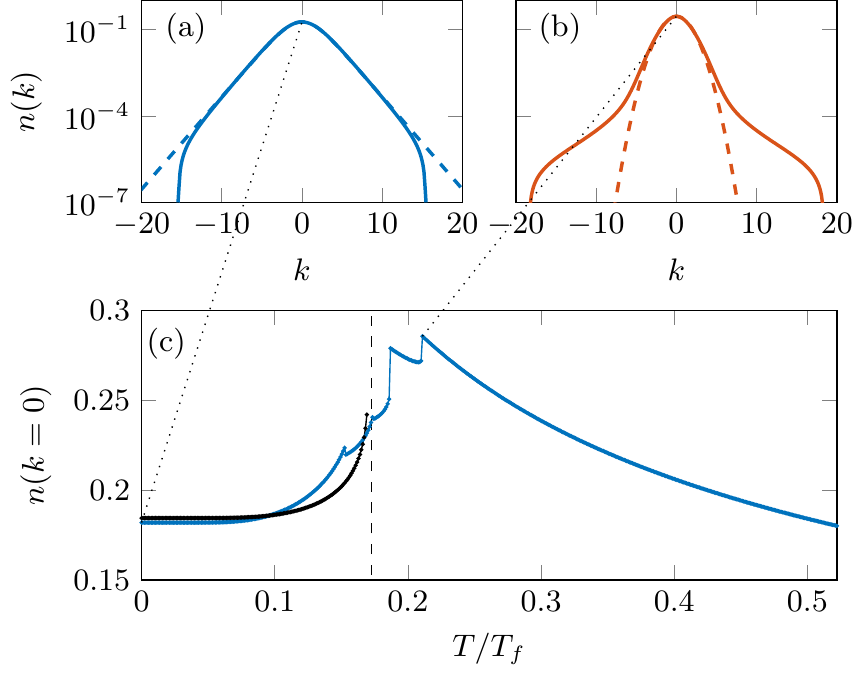}
    \caption{Momentum distribution of the TG gas for a system of $N=7$ particles and an interaction strength of $g_m=3$ at $T=0$ (a) and at $T=0.2109T_f$  (b). The blue dashed line shows the analytical curve from Eq. \eqref{eq:momdis_pinned} in the pinned state and the orange dashed line is a fitted Gaussian distribution with $\sigma = 1.3969$ in the thermal state respectively. (c) The zero-momentum component exhibits a peak around the transition point. Our model according to Eq.~\eqref{eq:momdis_pinned} (black dots), renormalized with $P_N$, agrees well with the numerical results (blue dots) below the critical temperature $T_\mathrm{crit} = 0.1726T_f$ (dashed line). Other parameters are $g_c=1$ and $\mu_0=200$.}
    \label{fig:momentum}
\end{figure}

Finally, direct observation of the transition between pinned and delocalized states can be made through the momentum distribution which can be obtained via time-of-flight measurements common to cold atom experiments. The momentum distribution can be calculated from the RSPDM as $n(k)=\int \rho(x,x') e^{-i k (x-x')} dx \, dx'$ and when the particles are pinned it also has a solitonic shape of inverse width $\pi/2a_\mathrm{pin}$
\begin{equation}
    n_\mathrm{pin}(k) = \frac{\pi}{4a_\mathrm{pin}}\frac{1}{\cosh^2\left(\frac{\pi}{2a_\mathrm{pin}}k\right)}\sum_{n=1}^{N}\frac{f_n}{N} \, .
    \label{eq:momdis_pinned}
\end{equation}
The numerically obtained momentum distribution of the TG gas in the pinned phase at $T=0$ is shown in Fig.~\ref{fig:momentum}(a) and it can be seen to agree well with the form of Eq.~\eqref{eq:momdis_pinned}. In the thermal phase (see Fig.~\ref{fig:momentum}(b)) the momentum distribution consists of a Gaussian peak for small momenta $k$ and additionally exhibits typical tails at large momenta for a TG gas at finite temperatures \cite{minguzzi2002high,olshanii2003short,paredes2004tonks,vignolo2013universal,xu2015universal}. Apart from the shape of the momentum distribution, the pinned and the thermal phase of the TG gas can also be distinguished by looking at the height of the zero-momentum peak of the normalized distribution, $n(k=0)$. As shown in Fig.~\ref{fig:momentum}(c), this value increases with increasing temperature in the pinned phase but decreases with increasing temperature in the thermal phase and is therefore maximal around the crossover. It also shows jumps whenever an individual particle is depinned.

\textit{Conclusions.} --- We have identified and characterized a self-pinning phase transition of a gas of strongly interacting bosons immersed in a Bose-Einstein condensate without any externally imposed lattice structure. The gas is pinned in a periodic manner once the interaction between the two components exceeds a certain value. We have presented a model to accurately describe the system in the pinned phase over a wide range of parameters and numerically calculated the phase diagram of the system.

We have also investigated the situation when the immersed component is at finite temperature and shown that the pinned state is unstable against thermal energies. Using a self-consistency criterion for the pinned state we have derived an expression for the critical temperature and shown that this behavior can be observed in the momentum distribution of the immersed atoms. Our work is a detailed investigation into a fundamental and complex many-body system that can be used to study new effects and is experimentally realizable. It also opens the door to studying quantum behavior in controllable environments (in this example made from a matter wave) and there are clear analogies to atoms confined in cavity fields \cite{Ritsch2013} and cold atom systems with long-range interactions \cite{Beau2020}. Including finite interactions between the immersed atoms would be an interesting future extension to this work, allowing one to explore how the competing intraspecies and interspecies interactions affect the phase diagram.

\begin{acknowledgments}
This work has been supported by the Okinawa Institute of Science and Technology Graduate University and used the computing resources of the Scientific Computing and Data Analysis section of the Research Support Division at OIST. We would like to thank Konrad Viebahn and the Lattice and Cavity Teams at ETH Zurich for inspiring discussions and their kind hospitality during a visit. T.K. acknowledges support from a Research Fellowship for Young Scientists by the Japan Society for the Promotion of Science under JSPS KAKENHI Grant No. 21J10521. T.F. acknowledges support from JSPS KAKENHI-21K13856.
\end{acknowledgments}



\section*{Supplementary material}

# Supplemental Material: Self-Pinning Transition of a Tonks-Girardeau Gas in a Bose-Einstein Condensate

Tim Keller, Thomás Fogarty, and Thomas Busch
*Quantum Systems Unit, Okinawa Institute of Science and Technology*
*Graduate University, Onna-son, Okinawa 904-0495, Japan*
(Dated: February 1, 2022)

## S1. FINITE SIZE EFFECTS

Since numerical studies of the self-pinning transition are restricted to a finite numbers of particles, we want to highlight some of the effects these finite system sizes have onto the numerical results and study the expected behavior in the limit $N \to \infty$. At $T = 0$ the Tonks-Girardeau (TG) gas exhibits a finite gap in its spectrum even in the noninteracting limit $g_m \to 0$ as seen in Fig. 1(c) in the main text. This is due to the finite system size and the value of this gap is determined by the box potential containing the TG gas according to

$$\Delta E(g_m = 0) = \frac{\pi^2}{2} \left(\frac{N}{L}\right)^2 \left(\frac{2}{N} + \frac{1}{N^2}\right). \tag{S1}$$

By fixing the density $N/L$ the gap vanishes in the limit $N \to \infty$ corresponding to the expected closing of the gap in the noninteracting case. This behavior is demonstrated in Fig. S1. At finite interaction strengths $g_m$ we find a universal behavior of the gap size independent of the number of particles in the TG gas above a sufficiently large system size of roughly $N \gtrsim 5$ as shown in the inset of Fig. S1. This is in accordance with our model where in the regime of a dilute TG gas the atoms in the pinned state can be regarded as individually localized impurities. As detailed in the paper, the energy gap roughly scales as $\Delta E \sim a_0^2/2 = g_m^4/8g_c^2$. Equating this with Eq. (S1) shows that the value of $g_m$ below which the gap closes, or saturates to its noninteracting value, scales as

$$g_m^* \sim \left[4\pi^2 g_c^2 \left(\frac{N}{L}\right)^2 \left(\frac{2}{N} + \frac{1}{N^2}\right)\right]^{\frac{1}{4}}. \tag{S2}$$

In the limit of $N \to \infty$ and fixed density this value vanishes as $g_m \sim N^{-1/4}$ meaning that, similarly to the pinning phase transition in an external lattice potential, almost arbitrarily small coupling strengths $g_m$ result in a pinned state [1].

At finite $T > 0$ the most prominent finite size effect is the extended range of temperatures over which the system gets gradually depinned, which influences both the phase diagram as well as the momentum distribution shown in the main text. In Fig. S2 we show the system's behavior for varying particle numbers at fixed density and interaction

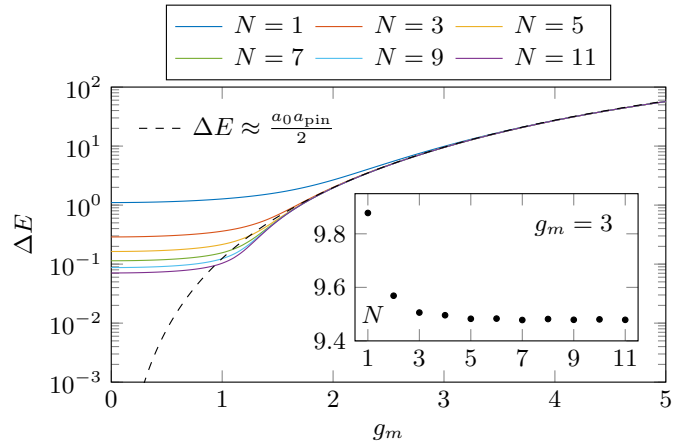

FIG. S1. Energy gap size in the pinned state as a function of interaction strength for varying particle numbers $N$ at $T = 0$ and fixed density $N/L = 11/40$. Other parameters are $g_c = 1$ and $\mu_0 = 200$. The inset shows a cross section for fixed $g_m = 3$. The dashed line shows an analytical approximation (see section S3).

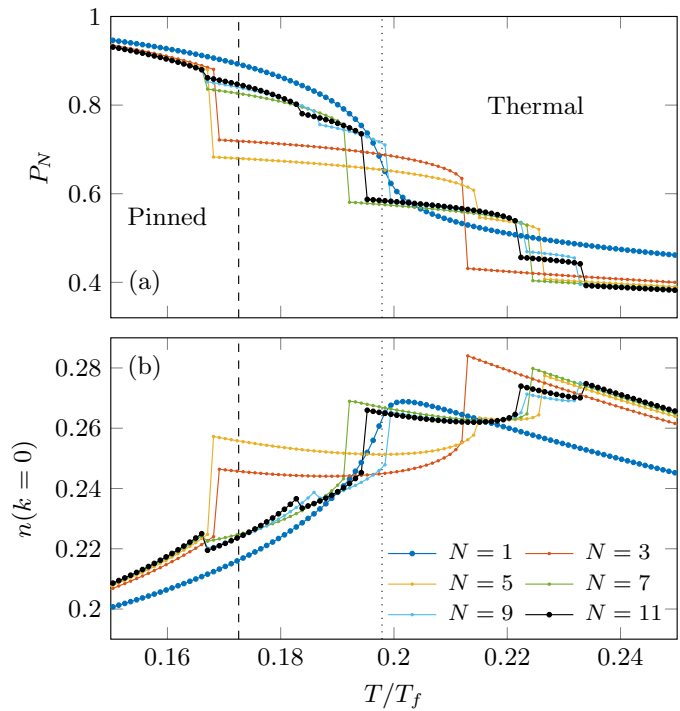

FIG. S2. (a) Ground state occupancy $P_N$ and (b) zero-momentum component of the momentum distribution $n(k)$ as a function of temperature around the critical region for varying particle numbers $N$ at fixed interaction strength $g_m = 3$ and density $N/L = 11/40$. Other parameters are $g_c = 1$ and $\mu_0 = 200$. The markers of the smallest and largest values $N = 1$ and $N = 11$ respectively are emphasized for clarity. The dashed line marks the analytical prediction $T_{\text{crit}} = 0.1726 T_f$ (see section S4) whereas the actual transition temperature is closer to the temperature $T = 0.1979 T_f$ marked by the dotted line where the ground state occupancy of the $N = 1$ lines reaches $P_N \approx 2/3$.

strength in the vicinity of the transition. In the limit of a single impurity $N = 1$ the transition between the pinned and thermal state is smooth in the ground state occupancy $P_N$ and the zero-momentum component of the momentum distribution $n(k = 0)$. The sudden jumps occurring for $N > 1$ are smoothed out with increasing system size while the temperature range over which they appear roughly remains the same. In the limit of $N \to \infty$ we expect the system to exhibit a single sudden jump in the vicinity of the $N = 1$ transition point at which the majority of particles is depinned simultaneously with otherwise smooth behavior before and after the transition.

## S2. ENERGY FUNCTIONAL

The atomic wave function in Eq. (6) in the main text does not take into account the energy needed by the impurity atoms to displace the Bose-Einstein condensate. Doing this leads to a reduction of the inverse width of the atomic wave function to $a < a_0 = g_m^2/2g_c$ and in order to determine this effect for the ground state we assume that $\phi$ in Eq. (6) in the main text can be written with a variable inverse width $a$ as

$$\phi(x) = \sqrt{\frac{a}{2}} \frac{1}{\cosh(ax)}. \tag{S3}$$

For the corresponding density dip in the BEC we assume an inverse width $b < a_0$, so that the condensate density can be written as

$$|\psi(x)|^2 = \frac{1}{g_c}\left(\tilde{\mu} - g_m \frac{b}{2}\frac{1}{\cosh^2(bx)}\right). \tag{S4}$$

These two free parameters can then be determined for the ground state in the pinned phase by minimizing the energy functional of the system as

$$E_{\text{tot}} = \min_{a,b}\left\{E_c^{\text{kin}}(b) + E_c^{\text{pot}}(b) + E_i^{\text{kin}}(a) + E_{\text{int}}(a,b)\right\}, \tag{S5}$$

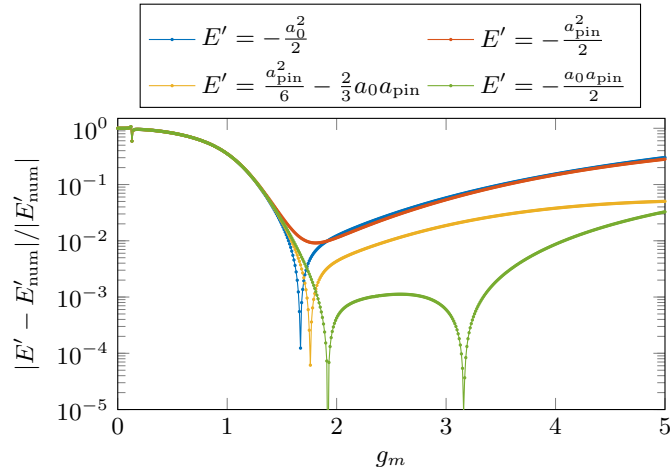

FIG. S3. Relative error for different approximations to the ground state energy of the effective Hamiltonian (S8) compared to the numerically obtained value of the ground state energy $E'_{\text{num}} = E_{1,\text{num}} - g_m \tilde{\mu}/g_c$ in the Tonks-Girardeau gases spectrum for the system of $N = 7$ particles as a function of $g_m$. Other parameters are $g_c = 1$ and $\mu_0 = 200$. See text for details.

where the individual terms are given by

$$E_c^{\text{kin}} = \frac{g_m^2 b^3}{8 g_c} \int_{-\infty}^{\infty} dy \frac{\sinh^2(y)}{\tilde{\mu} \cosh^6(y) - g_m \frac{b}{2} \cosh^4(y)} \tag{S6a}$$

$$\approx \frac{g_m^2}{30 \tilde{\mu} g_c} b^3,$$

$$E_c^{\text{pot}} = \frac{\tilde{\mu}^2 L_c}{2 g_c} - \tilde{\mu} \frac{g_m}{g_c} + \frac{g_m^2}{6 g_c} b, \tag{S6b}$$

$$E_i^{\text{kin}} = \frac{a^2}{6}, \tag{S6c}$$

$$E_{\text{int}} = \tilde{\mu} \frac{g_m}{g_c} - \frac{g_m^2}{4 g_c} \int_{-\infty}^{\infty} dx \frac{ab}{\cosh^2(ax) \cosh^2(bx)}. \tag{S6d}$$

The integral in Eq. (S6d) can be approximated by $\frac{8ab}{3(a+b)}$ if $a - b \ll 1$. For moderate values of $g_m$ it is sufficient to consider the case where $a = b = a_{\text{pin}}$, so that the system energy is minimized for

$$a_{\text{pin}} = \arg\min_{a} E_{\text{tot}}(a, a)$$

$$= -\frac{5\tilde{\mu}}{3} \frac{g_c}{g_m^2} + \sqrt{\frac{5\tilde{\mu}}{3} + \frac{25 \tilde{\mu}^2}{9} \frac{g_c^2}{g_m^4}}$$

$$= a_0 \frac{\sqrt{1 + 2\epsilon} - 1}{\epsilon}, \tag{S7}$$

with $\epsilon = 6 a_0^2 / 5 \tilde{\mu}$. Neglecting the kinetic energy of the BEC, i.e. letting $\epsilon \to 0$, the previous width $a_0 > a_{\text{pin}}$ is recovered.

## S3. SPECTRUM OF THE TONKS-GIRARDEAU GAS

The effective Hamiltonian for a single immersed impurity

$$\hat{H}'_1 = -\frac{1}{2} \frac{\partial^2}{\partial x^2} - 2 a_0 |\phi_1(x)|^2 \tag{S8}$$

does not capture the reduction in the peak height of the pinned wave function due to $a_{\text{pin}} < a_0$. Among other factors, this reduction leads to a difference between the numerically observed spectrum of the Tonks-Girardeau gas

in the pinned state and the analytical ground state value of the effective Hamiltonian given by $E' = -a_0^2/2$ for $\phi_1(x) = \sqrt{a_0/2}\cosh^{-1}(a_0 x)$ (see Fig. S3). Simply replacing $a_0$ by $a_{\text{pin}}$ both in the effective Hamiltonian as well as in $\phi_1(x)$, yielding $E' = -a_{\text{pin}}^2/2$, leads to a similar relative error but this time underestimating the magnitude of $E'$. The closer approximation which we are deriving and using in the manuscript above is given by the expectation value of the original effective Hamiltonian with the numerically observed state $\phi_1(x) = \sqrt{a_{\text{pin}}/2}\cosh^{-1}(a_{\text{pin}}x)$ according to

$$E' \approx \langle \phi_1 | \hat{H}_1' | \phi_1 \rangle = \frac{a_{\text{pin}}^2}{6} - \frac{2}{3}a_0 a_{\text{pin}}. \tag{S9}$$

However it is worth noting that we heuristically found that an even better approximation is given by $E' = -a_0 a_{\text{pin}}/2$ (see Fig. S3), which corresponds to replacing one of the factors of $a_{\text{pin}}$ in the first term in Eq. (S9) by $a_0$.

### S4. CRITICAL TEMPERATURE

If the Tonks-Girardeau gas has a finite temperature, its ground state is determined by the self-consistency equation

$$f_{\text{pin}} = \frac{1}{\exp\left\{\frac{1}{T}\left[E(f_{\text{pin}}) - \mu(f_{\text{pin}})\right]\right\} + 1}, \tag{S10}$$

where we have set $k_{\text{B}} = 1$ for simplicity. In order to find the critical temperature at which the system undergoes a transition between the pinned and a thermal state we first calculate the derivative of $f_{\text{pin}}$ with respect to $T$

$$\begin{aligned}\frac{\partial f_{\text{pin}}}{\partial T} &= -\frac{f_{\text{pin}}^2 - f_{\text{pin}}}{T^2}\left[E(f_{\text{pin}}) - \mu(f_{\text{pin}}, T) + T\frac{\partial \mu}{\partial T}\right] \\ &+ \frac{f_{\text{pin}}^2 - f_{\text{pin}}}{T}\left(\frac{\partial E}{\partial f_{\text{pin}}} - \frac{\partial \mu}{\partial f_{\text{pin}}}\right)\frac{\partial f_{\text{pin}}}{\partial T}.\end{aligned} \tag{S11}$$

This shows that at a fixed point of the self-consistency equation with $0 < f_{\text{pin}} < 1$ the system additionally needs to fulfill

$$E(f_{\text{pin}}) - \mu(f_{\text{pin}}, T) + T\frac{\partial \mu}{\partial T} = 0, \tag{S12a}$$

$$T = \left(f_{\text{pin}}^2 - f_{\text{pin}}\right)\left(\frac{\partial E}{\partial f_{\text{pin}}} - \frac{\partial \mu}{\partial f_{\text{pin}}}\right). \tag{S12b}$$

Eq. (S12b) also describes the point at which both sides of the self-consistency equation (S10) are tangent to each other at the fixed point and therefore have identical derivatives with respect to $f_{\text{pin}}$.

The temperature dependence of the chemical potential can be approximated by a Sommerfeld expansion [2]

$$\begin{aligned}\mu(f_{\text{pin}}, T) &\approx \epsilon_F - \frac{\pi^2}{6}T^2\frac{g'(\epsilon_F)}{g(\epsilon_F)} \\ &= \frac{E(f_{\text{pin}})}{2} + \frac{\pi^2}{6}\frac{T^2}{E(f_{\text{pin}})},\end{aligned} \tag{S13}$$

where we assumed that the density of states still obeys $g(\epsilon_F) \sim 1/\sqrt{\epsilon_F}$ for our one-dimensional box potential and where we chose to center the Fermi level in the middle of the gap according to $\epsilon_F = E(f_{\text{pin}})/2$. We note that the Fermi level is not constant but implicitly depends on $T$ via $f_{\text{pin}}$ as the energy gap decreases for increasing temperature. For $E(f_{\text{pin}}) = E'(g_m \to g_m\sqrt{f_{\text{pin}}})$ we use the heuristically found approximation

$$E(f_{\text{pin}}) \approx -\frac{1}{2}f_{\text{pin}}a_0 a_{\text{pin}} = -a_0^2\frac{\sqrt{1+2\epsilon f_{\text{pin}}^2}-1}{2\epsilon}, \tag{S14}$$

in order to obtain a best possible estimate for the critical temperature. Using these approximations in the right-hand side of Eq. (S12b) and neglecting the weak dependence of $\tilde{\mu}$ on $f_{\text{pin}}$ within $\epsilon$ gives

$$\begin{aligned}T &= \left(f_{\text{pin}}^2 - f_{\text{pin}}\right)\left(\frac{\partial E}{\partial f_{\text{pin}}} - \frac{\partial \mu}{\partial f_{\text{pin}}}\right) \\ &\approx \frac{f_{\text{pin}}^2 - f_{\text{pin}}^3}{\sqrt{1+2\varepsilon f_{\text{pin}}^2}}a_0^2\left[\frac{1}{2} + \frac{\pi^2}{6}\frac{T^2}{E(f_{\text{pin}})^2}\right].\end{aligned} \tag{S15}$$

Since $f_{\text{pin}}$ decreases as a function of $T$ this indicates that the fixed point changes from a value in the pinned phase to a value in the thermal phase once its value drops below

$$f_{\text{pin}} < f^* = \max_{0 \leq f \leq 1} \left\{ \frac{f^2 - f^3}{\sqrt{1 + 2\epsilon f^2}} \right\} \approx \frac{2}{3}, \tag{S16}$$

where the first factor in the previous equation (S15) has a maximum as a function of $f_{\text{pin}}$. Finally, using again the Sommerfeld expansion for the chemical potential we fix $f_{\text{pin}} = f^*$ in the self-consistency equation (S10) and solve for $T$ which we define as the critical temperature. This yields

$$T_{\text{crit}} = E'(f^*) \frac{3 \ln(\frac{1}{f^*} - 1)}{\pi^2} \left[ \sqrt{1 + \frac{\pi^2}{3 \ln^2(\frac{1}{f^*} - 1)}} - 1 \right]$$
$$\equiv C(f^*) E'(f^*), \tag{S17}$$

or in terms of $T_f$

$$\frac{T_{\text{crit}}}{T_f} = C(f^*) \frac{\sqrt{1 + 2(f^*)^2 \epsilon} - 1}{\sqrt{1 + 2\epsilon} - 1}$$
$$\approx 0.3795 \frac{\sqrt{1 + \frac{8}{9}\epsilon} - 1}{\sqrt{1 + 2\epsilon} - 1} \quad \text{for} \quad f^* \approx \frac{2}{3}. \tag{S18}$$

---


\end{document}